\documentclass{PoS}
\usepackage[numbers,sort&compress]{natbib}
\usepackage{natbibspacing}
\title{The $B\to \pi l \nu$ form factor from unquenched lattice QCD with domain-wall light quarks
and relativistic $b$-quarks}

\ShortTitle{The $B\to \pi l \nu$ form factor}

\author{\speaker{Taichi Kawanai}\\
Physics Department, Brookhaven National Laboratory, Upton, NY 11973, USA\\
RIKEN-BNL Research Center, Brookhaven National Laboratory, Upton, NY 11973, USA\\
Department of Physics, The University of Tokyo, Hongo 7-3-1, Tokyo 113-0033, Japan\\
        E-mail: \email{kawanai@nt.phys.s.u-tokyo.ac.jp}}

\author{Ruth S.~Van de Water\\
         Physics Department, Brookhaven National Laboratory, Upton, NY 11973, USA\\
	Theoretical Physics Department, Fermi National Accelerator Laboratory, Batavia, IL  60510, USA\\
        E-mail: \email{ruthv@fnal.gov}}

\author{Oliver Witzel\\
        Center for Computational Science, Boston University, 3 Cummington Mall, Boston, MA 02215, USA\\
	E-mail: \email{owitzel@bu.edu}
      }

\abstract{We report on  a lattice-QCD calculation of the $B \to \pi l\nu$ form factor  with 
domain-wall light quarks and relativistic $b$-quarks
using the $2+1$ flavor domain-wall fermion and Iwasaki gauge-field ensembles
generated by the RBC and UKQCD Collaborations. 
We present initial results  obtained 
from the coarser ($a\approx 0.11$~fm) $24^3$ lattices  
and some of the finer ($a\approx 0.086$~fm) $32^3$ lattices.
}

\FullConference{The 30 International Symposium on Lattice Field Theory - Lattice 2012,\\
		June 24-29, 2012\\
		Cairns, Australia}

\begin{document}
\section{Introduction}
The theoretical  calculation of the hadronic $B\to\pi l\nu$ form factor $f_+(q^2)$
is a key ingredient in the determination of the Cabibbo-Kobayashi-Maskawa~(CKM) 
matrix element $|V_{ub}|$
from  $B \to \pi l\nu$ exclusive semi-leptonic decay.
The value of $|V_{ub}|$ characterizes the strength of the quark-flavor changing $b\to u$ transition,
and can be  obtained by combining the hadronic $B\to\pi l\nu$ form factor $f_+(q^2)$ 
with experimental measurements of the differential decay rate via 
\begin{equation}
 \frac{d\Gamma(B\to \pi l\nu)}{dq^2} = 
\frac{G_F^2|V_{ub}|^2}{192\pi^2m_B^3}\left[
  (m_B^2+m_\pi^2-q^2)^2-4m_B^2m_\pi^2
\right]^{3/2} |f_+(q^2)|^2,
\end{equation}
where the momentum transfer $q^\mu\equiv p_B^\mu - p_\pi^\mu$ and
 we neglect the mass of the outgoing lepton.
The form factor $f_+(q^2)$ encodes nonperturbative QCD dynamics, 
and can only be computed precisely from first principles using lattice QCD.
Indeed there have been two $2+1$ flavor lattice calculations of $f_+(q^2)$
done by the HPQCD~\cite{Dalgic:2006dt} and FNAL/MILC Collaborations~\cite{Bailey:2008wp}.
Both groups use the MILC gauge configurations.

The precise calculation of $|V_{ub}|$ constrains 
the apex of the CKM unitarity triangle.
There is a persistent puzzle, however,  between independent determinations of $|V_{ub}|$.
First, the exclusive determination from $B \to \pi l\nu$ and 
the inclusive determination from $B \to X_u l\nu$ 
(where $X_u$ is any charmless hadronic final state)
differ at the level of 
more than $3\sigma$~\cite{Antonelli:2009ws}.
Second, most experimental measurements of BR ($B\to\tau\nu$) combined with lattice-QCD input for $f_B$
give a  higher value of $|V_{ub}|$ 
than both $|V_{ub}|_{\rm excl}$ and $|V_{ub}|_{\rm incl}$~\cite{Lunghi:2010gv,Laiho:2012ss},
although the recent Belle measurement of  BR ($B\to\tau\nu$)  is 
lower and more in line with the Standard-Model prediction, albeit still with large errors~\cite{Adachi:2012mm}.
Given this situation, an independent lattice calculation using a a different gauge action 
is desired  to address this puzzle.

\section{Methodology}
The $B\to \pi l \nu$ hadronic weak matrix element is parameterized by 
the  form factors $f_+(q^2)$ and $f_0(q^2)$ as 
\begin{equation}
 \langle  \pi | \mathcal{V}^\mu | B \rangle 
= f_+(q^2) \left( p_B^\mu + p_\pi^\mu - \frac{m_B^2-m_\pi^2}{q^2}q^\mu\right)
f_0(q^2)\frac{m_B^2-m_\pi^2}{q^2}q^\mu,
\end{equation}
where the $b\to u$ vector current $\mathcal{V}^\mu\equiv i\bar{u}\gamma^\mu b$. 
We calculate numerically the form factors $f_{\parallel}$ and $ f_{\perp}$,
which are more convenient for the lattice calculation:
\begin{equation}
 \langle  \pi | \mathcal{V}^\mu | B \rangle = \sqrt{2m_B}
[v^\mu f_{\parallel}(E_\pi) + p^\mu_{\perp}f_{\perp}(E_\pi)],
\end{equation}
where $v^\mu = p_B^\mu/m_B$ and $p^\mu_{\perp}=p_\pi^\mu -(p_\pi\cdot v)v^\mu$.
In the $B$-meson rest frame, these form factors are
 proportional to the hadronic matrix elements of the temporal and spatial vector current:
\begin{equation}
f_{\parallel} = \frac{\langle \pi | \mathcal{V}^0 |B\rangle }{\sqrt{2m_B}} , \ \ \ \ 
 f_{\perp} = \frac{\langle \pi | \mathcal{V}^i | B\rangle}{\sqrt{2m_B}} \frac{1}{p^i_\pi}. 
\end{equation}
The desired form factor $f_+$ can be obtained by following relation:
\begin{equation}
 f_+(q^2) = \frac{1}{\sqrt{2m_B}}[f_{\parallel}(E_\pi)+(m_B-E_\pi)f_{\perp}(E_\pi)].
\end{equation}

We match the lattice amplitude to the continuum matrix element
using the mostly nonperturbative method of Ref.~\cite{ElKhadra:2001rv}:
\begin{equation}
\langle \pi | \mathcal{V}^\mu | B\rangle = Z_{V_\mu}^{bl} \langle \pi | V^\mu | B\rangle,
 \ \ \ \ 
Z_{V_\mu}^{bl} = \rho_{V_\mu}^{bl} \sqrt{Z_{V}^{bb} Z_{V}^{ll} }.
\end{equation}
The flavor-conserving renormalization factors $Z_{V}^{bb}$ and  $ Z_{V}^{ll}$  are 
computed nonperturbatively 
on the lattice and the factor $\rho$ is computed at one loop in mean-field improved
lattice perturbation theory~\cite{Lepage:1992xa}.
Most of the heavy-light current renormalization factor comes from $Z_V^{bb}$ and $Z_V^{bl}$, 
such that $\rho$ is expected to be close to unity~\cite{Harada:2001fi}.

We improve the $b \to u$ vector current through $\mathcal{O}(\alpha_S a)$.  
At this order we need only compute one additional matrix element with a single-derivative operator.  
We calculate the improvement coefficient at 1-loop in mean-field improved lattice perturbation theory.  
We have not yet included the $\mathcal{O}(a)$-improvement term in the results shown in these proceedings.

\section{Computational setup}
\begin{table}[t]
  \centering
  \caption{ Lattice simulation parameters.
    The results reported here use the ensembles specified in boldface
}
  \label{tab:ensembles}
 \begin{tabular}{ccccccc} \hline\hline 
  $a$ [fm] & $L^3\times T$ & $am_l$ & $am_s$ & $M_\pi$ [MeV] & \# configs. & \# time sources \\ \hline
  $\mathbf{\approx 0.11}$ & $\mathbf{24^3\times 64}$ & \bf 0.005 & \bf 0.040 & \bf 329 & \bf 1636 & \bf 1 \\
  $\mathbf{\approx 0.11}$ & $\mathbf{24^3\times 64}$ & \bf 0.010 & \bf 0.040 & \bf 422 & \bf 1419 & \bf 1 \\ \hline
  $\mathbf{\approx 0.086}$& $\mathbf{32^3\times 64}$ & \bf 0.004 & \bf 0.030 & \bf 289 & \bf 628 & \bf 2 \\
  $\approx 0.086$ & $32^3\times 64$ & 0.006 & 0.030 & 345 & 889 & 2 \\ 
  $\approx 0.086$ & $32^3\times 64$ & 0.008 & 0.030 & 394 & 544 & 2 \\ \hline\hline

 \end{tabular}
\end{table}
\begin{figure}[t] 
  \centering
  \includegraphics[width=.70\textwidth]{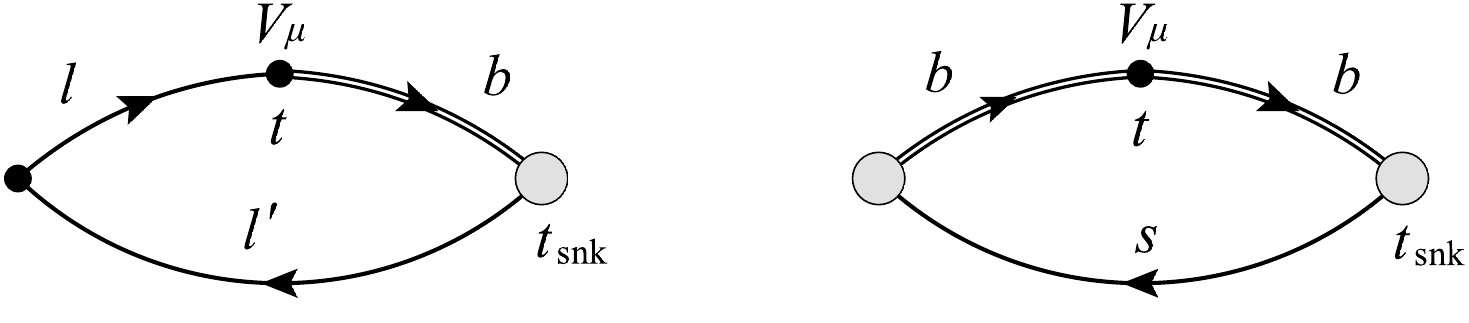}
  \caption{Three-point correlation functions for computing the $B\to \pi l\nu$ form factor (left) and 
  renormalization factor $Z_V^{bb}$ (right). 
  The single and double lines correspond to light-  and $b$-quark propagators, respectively. 
   The spectator light quark is labeled $l'$ and the daughter light quark is labeled $l$.
  Shaded cricles denote the  gauge invariant Gaussian smeared source/sink for the 
  $b$-quarks.
 }
   \label{fig:3pt_diagram}
\end{figure}
Our computation of the $B\to\pi l\nu$ form factor is performed on 
$2+1$ flavor domain-wall fermion and Iwasaki gauge-field ensembles
generated by the RBC and UKQCD Collaborations
with several values of the light dynamical quark mass 
at two lattice spacings, the coarser $a\approx 0.11$~fm ($a^{-1}\approx 1.73$~GeV) 
and the finer $a\approx 0.086$~fm ($a^{-1}\approx 2.28$~GeV)~\cite{Allton:2008pn,Aoki:2010dy}.
The set of  gauge-field configurations used for this project  is summarized in Table~\ref{tab:ensembles}.
The results reported here use the ensembles specified in boldface.
On the finer ensembles we compute two quark propagators on each configuration 
with their temporal source locations separated by $T/2$ to increase the statistics.
Periodic boundary conditions are imposed in the time direction.
In this proceedings, we present first results obtained 
from the coarser ensembles and a subset of the finer ensembles.

For the light quarks, domain-wall valence quark propagators are generated 
on each configurations with several partially quenched masses
 in order to enable good control over the chiral extrapolation.
For the bottom quark, we use the relativistic heavy quark (RHQ) action~\cite{Christ:2006us}
to remove the large discretization error introduced by the large bottom quark mass~\cite{ElKhadra:1996mp}.
The RHQ action is given by
\begin{equation}
 S_{\rm RHQ} = \sum_{n,m} \bar{q}_n \left\{
m_0+\gamma_0D_0- \frac{aD_0^2}{2} + \xi\left[ \vec{\gamma}\cdot\vec{D} - 
\frac{a(\vec{D})^2}{2}\right]
 - a\sum_{\mu,\nu} \frac{ic_P}{4}\sigma_{\mu\nu}F_{\mu\nu}
\right\}_{n, m} q_m
\end{equation}
where tuning the bare-quark mass $m_0 a$, the clover coefficient $c_P$, and the anisotropy parameter $\xi$
for the $b$ quark
is needed~\cite{Christ:2006us, Lin:2006ur}.
Here we employ values determined nonperturbatively  in Ref.~\cite{Aoki:2012xaa}.

Figure~\ref{fig:3pt_diagram} shows the the three-point correlation functions needed in this project.
The left diagram of Fig.~\ref{fig:3pt_diagram} depicts the computation of the $B \to \pi l\nu$ form factor.
The source of the pion (with both zero and nonzero momenta) is located at the origin.
The $B$-meson is at rest at $t_{\rm snk}$.
We compute this three-point function with a unitary spectator mass and 
multiple partially quenched daughter-quark masses
that enable interpolation to the physical strange-quark mass 
and extrapolation to the physical average up-down quark mass.
The right diagram of Fig.~\ref{fig:3pt_diagram} is used to obtain the  renormalization factor $Z_V^{bb}$.
We fix the the spectator mass to be $m_q$ = $m_s$ in order to reduce the statistical errors
because $Z_V^{bb}$ is independent of the light spectator-quark mass.
For $Z_V^{ll}$, we use the value obtained by the RBC/UKQCD Collaborations
in Ref.~\cite{Aoki:2010dy}.
 To reduce excited-state contamination,
 we employ a gauge-invariant Gaussian smeared (sequential) source 
for the $B$-meson in all lattice correlators.
In order to reduce the statistical error of the three point function,
we calculate the three point function with the $B$-meson
 at $t_{\rm snk}$ and $(T-t_{\rm snk})$ and average the results.

\section{Two-point and three-point fits}
\begin{figure}[t] 
  \centering
  \includegraphics[width=.49\textwidth]{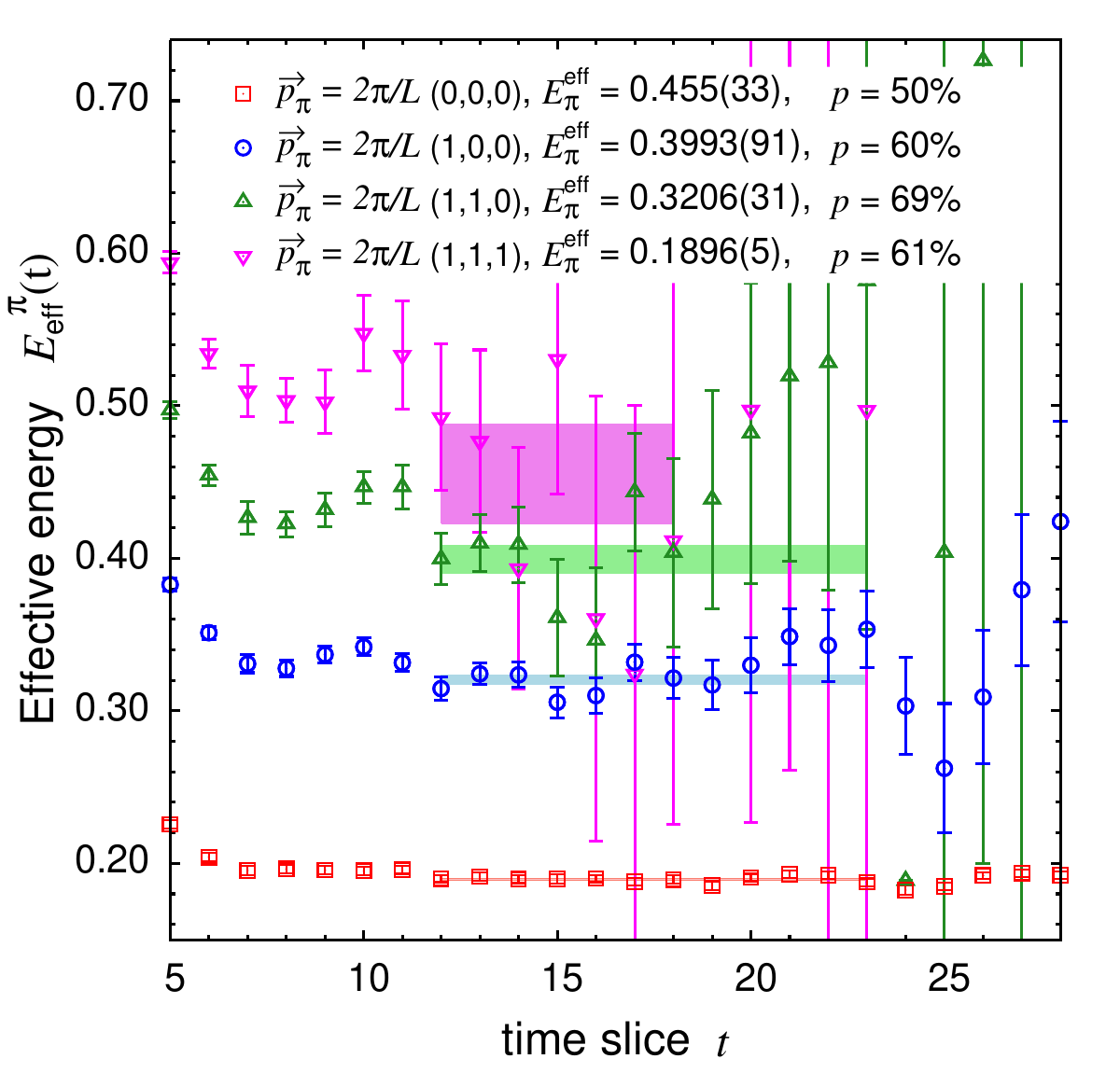}
  \includegraphics[width=.49\textwidth]{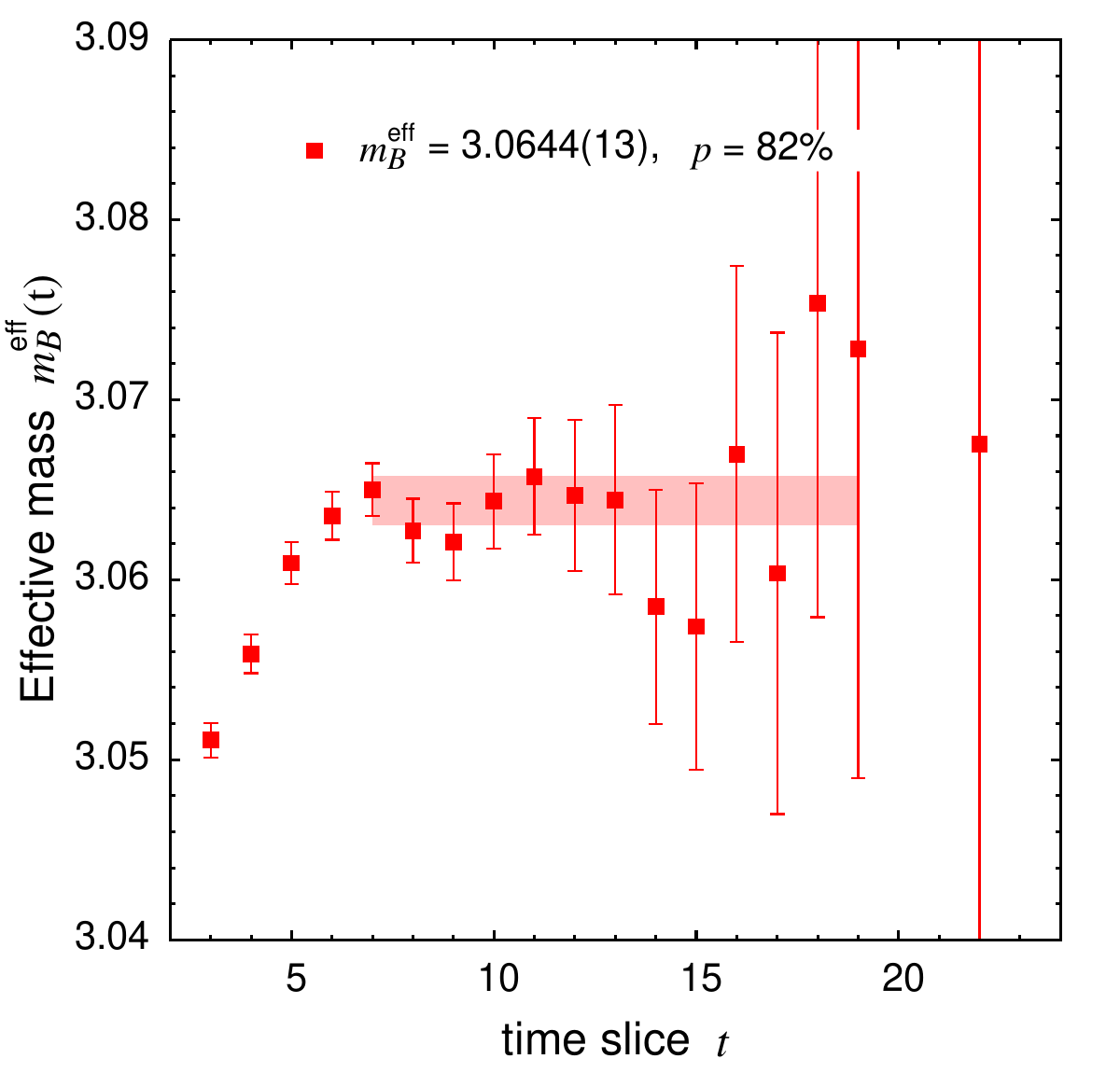}
  \caption{ Effective energy/mass plots for the  
    pion~(left) and $B$-meson~(right) on the coarser $am_l = 0.005$  ensemble.
    Shaded bands show the fit results with jackknife statistical errors and fit ranges.
    For the pion, data points for four spatial momenta $(\vec{p}L/2\pi)^2 = 0, 1, 2, 3$  
     are indicated by different colors/symbols.}
   \label{fig:2pt}
\end{figure}
The pion energy and $B$-meson mass are obtained from the following two point functions:
\begin{eqnarray}
    C_2^{\pi}(t,\vec{p}) &=&  \sum_{\vec{x}} e^{i\vec{p}_\pi\cdot \vec{x}} 
        \langle \mathcal{O}_\pi(t,\vec{x}) \mathcal{O}_\pi^\dagger(0,\vec{0}) \rangle, \\
    C_2^{B}(t) &=&  \sum_{\vec{x}}  
        \langle \mathcal{O}_B(t,\vec{x}) \mathcal{O}_B^\dagger(0,\vec{0}) \rangle,
\end{eqnarray}
where $\mathcal{O}_\pi$ and $\mathcal{O}_B$ are interpolating operators for the pion 
and $B$-meson.
We compute the effective energies as
\begin{equation}
 E_{\rm eff}(\vec{p})= \cosh^{-1}
\left[
  \frac{C_2(t,\vec{p})+C_2(t+2,\vec{p})}{C_2(t+1,\vec{p})}
\right].
\end{equation}
The left plot of Fig.~\ref{fig:2pt} shows the effective 
pion energies for momenta through  $(\vec{p}L/2\pi)^2 = 3$ on the  the coarser 
$am_l = 0.005$ ensemble.
Similarly, the right plot of Fig.~\ref{fig:2pt} shows the effective $B$-meson mass
calculated with a Gaussian smeared source and point sink.

\begin{figure}[t] 
  \centering
  \includegraphics[width=.49\textwidth]{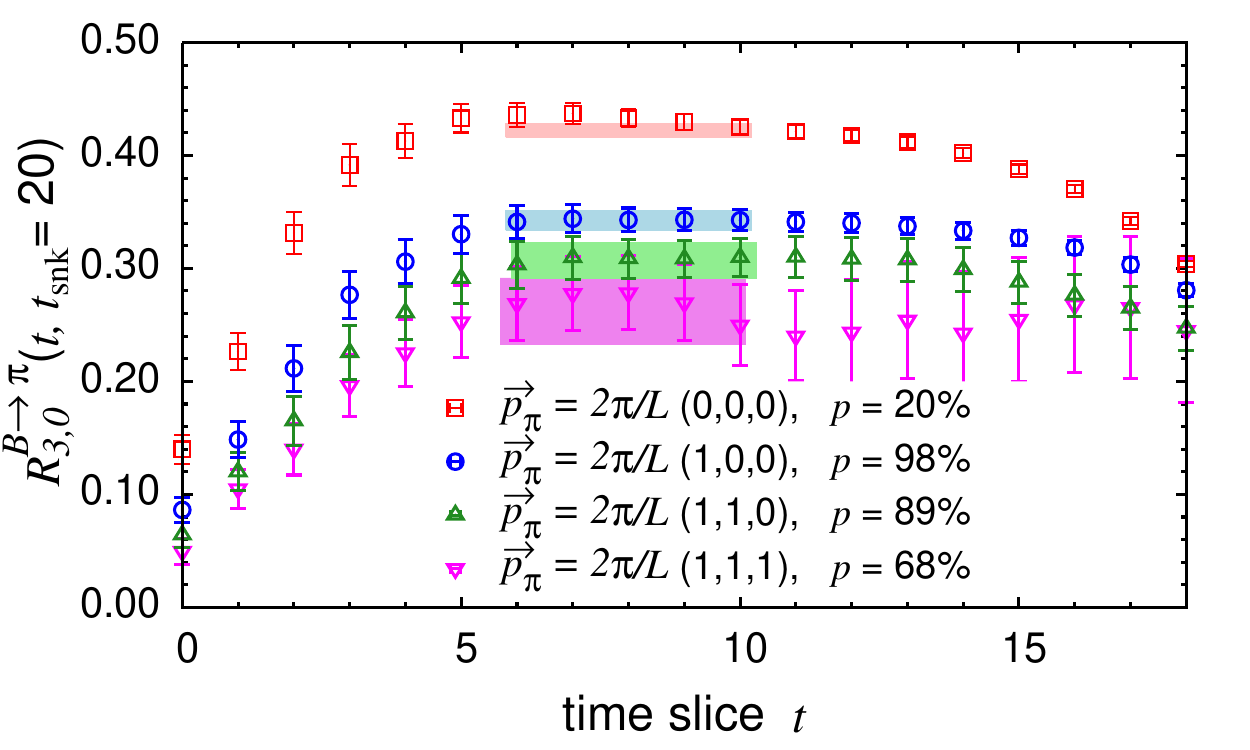}
  \includegraphics[width=.49\textwidth]{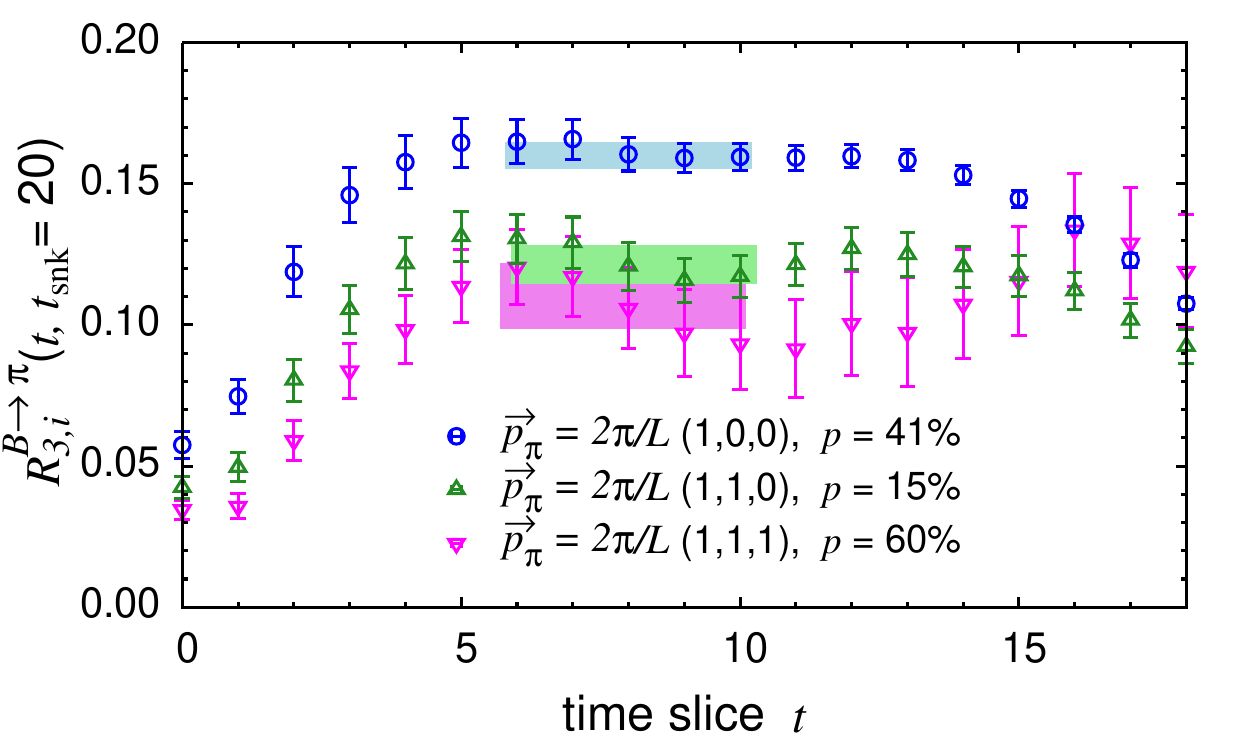}
  \caption{The ratio $R_{3,0}^{B\to \pi}$ (left) and $R_{3,i}^{B\to \pi}$ (right)
  with $t_{\rm snk} = 20$ on the  coarser $am_l = 0.005$ ensemble.
  Fit ranges and fit results with jackknife statistical errors are shown as horizontal bands.}
   \label{fig:3pt}
\end{figure}
In the lattice-QCD simulations, 
the form factors $f_{\parallel}^{\rm lat }$ and $ f_{\perp}^{\rm lat }$  
can be obtained from the following 
ratios of correlation functions:
\begin{eqnarray}
R_{3,\mu}^{B\to \pi}(t,t_{\rm snk}) &=& \frac{C_{3,\mu}^{B\to \pi}(t,t_{\rm snk})}{\sqrt{C_2^{\pi}\
(t) C_2^{B}(t_{\rm snk}-t) }}
     \sqrt{\frac{2E_\pi}{e^{-E_\pi t}e^{-m_B t}}} \\
     f_{\parallel}^{\rm lat }&=&\lim_{0\ll t \ll t_{\rm snk}}R_0^{B\to \pi}(t,t_{\rm snk}), \\
     f_{\perp}^{\rm lat }&=&\lim_{ 0\ll t \ll t_{\rm snk}}\frac{1}{p^i_\pi}R_i^{B\to \pi}(t,t_{\rm snk}),
\end{eqnarray}
where 
\begin{equation}
    C_{3,\mu}^{B\to \pi}(t,t_{\rm snk}, \vec{p}) =  \sum_{\vec{x},\vec{y}} e^{i\vec{p}_\pi\cdot \vec{y}} 
        \langle \mathcal{O}_B(t_{\rm snk},\vec{x}) V_\mu(t,\vec{y})
	  \mathcal{O}_\pi^\dagger(0,\vec{0}) \rangle.
\end{equation}
$V_\mu$ is the vector current matrix element on the lattice.

Fig.~\ref{fig:3pt} shows the results for  $R_{3,0}^{B\to \pi}$ and $R_{3,i}^{B\to \pi}$
calculated with a source-sink separation $t_{\rm snk} = 20$ on the  coarser  $am_l = 0.005$  ensemble.
For  $R_{3,i}^{B\to \pi}$, we average over equivalent spatial momenta.
Plots of these ratios for the other ensembles and partially quenched masses look similar.
We determine $f_{\parallel}^{\rm lat }$ and $ f_{\perp}^{\rm lat }$   
by fitting to a plateau in the region $0\ll t \ll t_{\rm snk}$
where  we expect the excited-state contributions to be negligible 
and obtain a good correlated $\chi^2/{\rm d.o.f.}$ and $p$-value.

\section{Renormalization factor $Z_V^{bb}$}
\begin{figure}[t] 
  \centering
  \includegraphics[width=.50\textwidth]{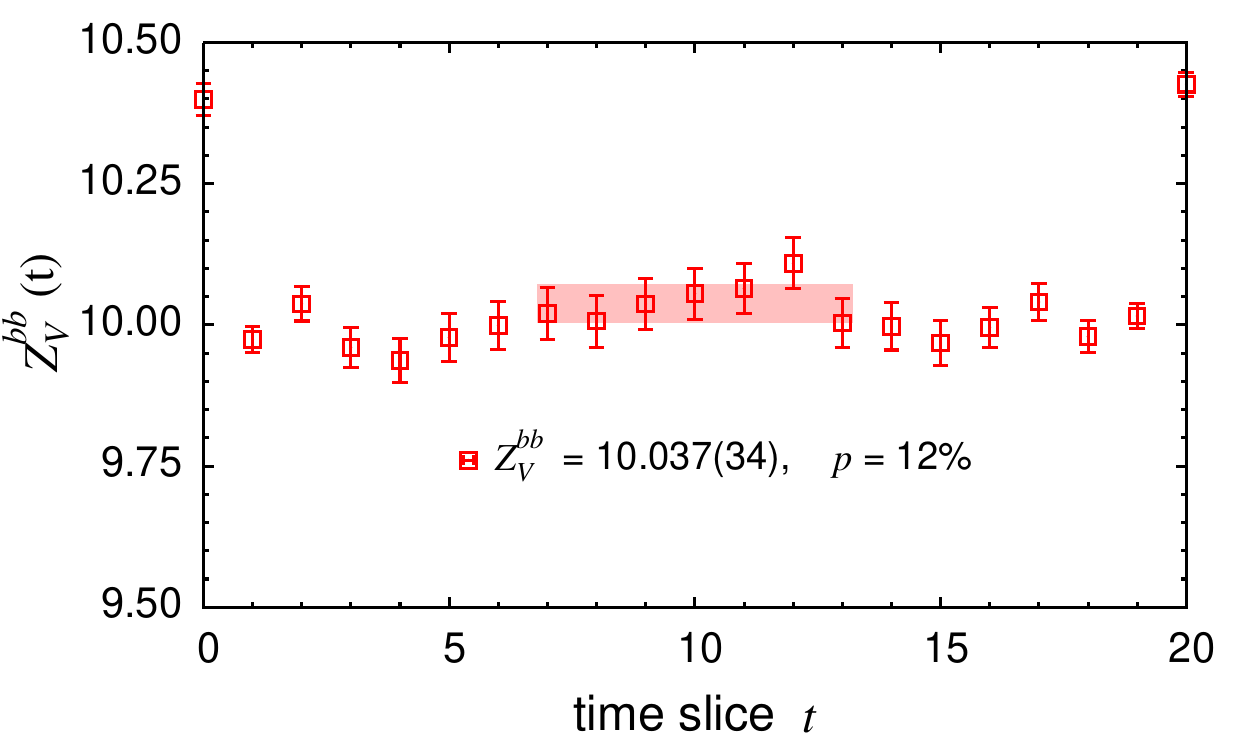}
  \caption{Determination of the renormalization factor  $Z_V^{bb}$. 
    with $t_{\rm snk} = 20$ on the  coarser  $am_l = 0.005$ ensemble.
  The shaded band shows the fit result with jackknife statistical error.}
   \label{fig:form}
\end{figure}
 Values of $Z_V^{bb}$ can be  computed 
nonperturbatively using the charge-normalization condition
$Z_V^{bb}\langle B_s|V^{bb,0}|B_s\rangle =1$
where $V^{bb,0}$ is the $b\to b$ lattice vector current.
We obtain $Z_V^{bb}$ by fitting the following ratio of correlators
 to a plateau in a region $0 \ll t \ll t_{\rm snk}$:
\begin{equation}
 Z_V^{bb}(t,t_{\rm snk}) = C_2^{B_s}(t_{\rm snk})\Big/C_{3,0}^{B_s\to B_s}(t,t_{\rm snk}).
\end{equation}
The determination of  $Z_V^{bb}$ on the  coarser  $am_l = 0.005$ ensemble
is shown in Fig.~\ref{fig:form}.
At tree level in mean-field improved lattice perturbation theory,
 $Z_V^{bb}$ is given by~\cite{Christ:2006us,ElKhadra:1996mp}
\begin{equation}
 Z_V^{bb} = u_0 \exp(M_1), \ \ M_1=\log\left[1+\tilde{m}_0\right], \ \ 
\tilde{m}_0 = \frac{m_0}{u_0} -(1+3\zeta)(1-\frac{1}{u_0}).
\end{equation}
Using $m_0 = 8.45$, $\zeta = 3.1$ and $u_0 = 0.8757$ for the coarser $am_l = 0.005$ ensemble
\cite{Aoki:2010dy,Aoki:2012xaa}, we
obtain a tree level estimate $Z_V^{bb}= 10.606$, which is 
in good agreement with the nonperturbative determination.

\section{Form-factor results and outlook}
\begin{figure}[t] 
  \centering
  \includegraphics[width=.49\textwidth]{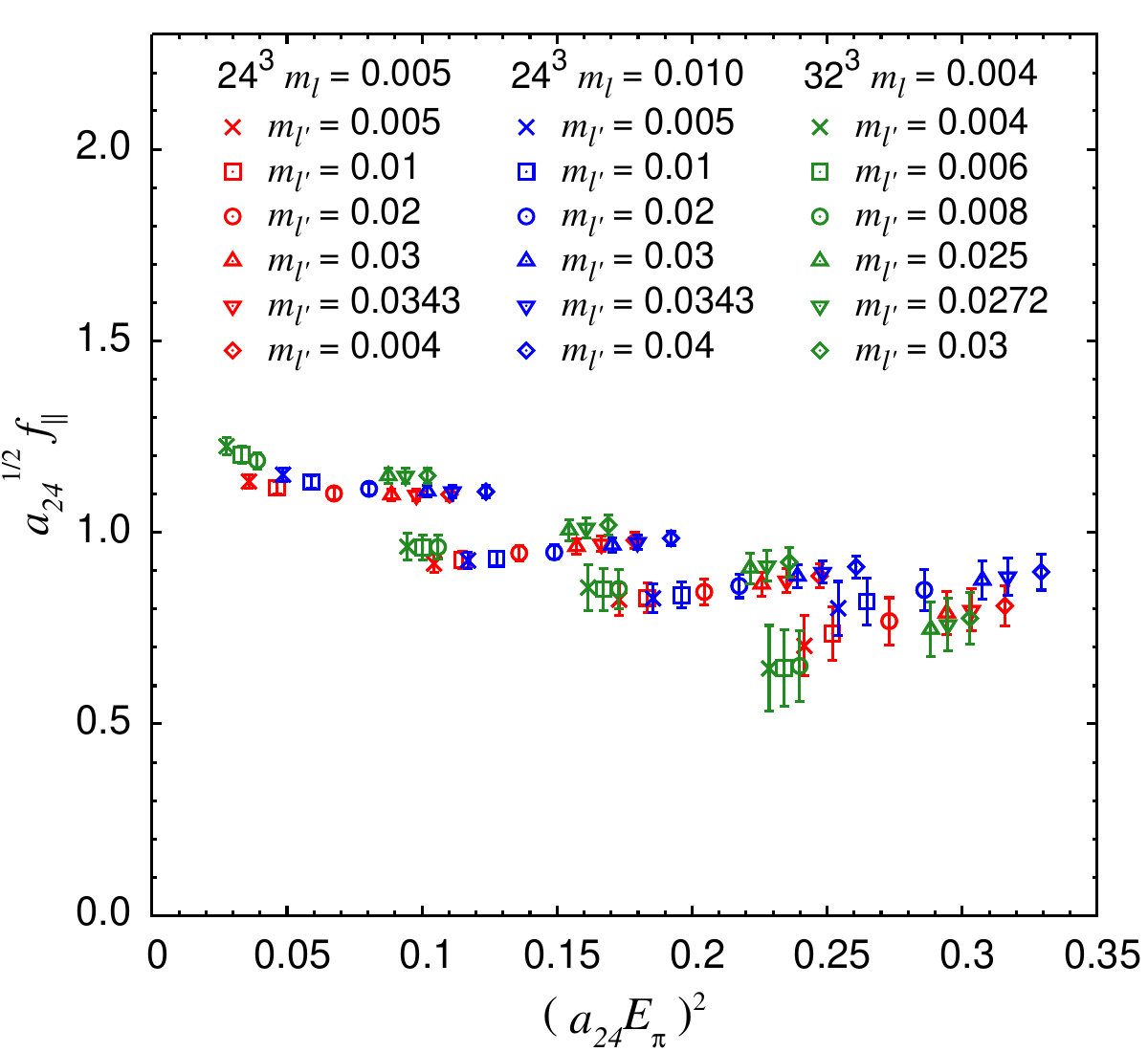}
  \includegraphics[width=.49\textwidth]{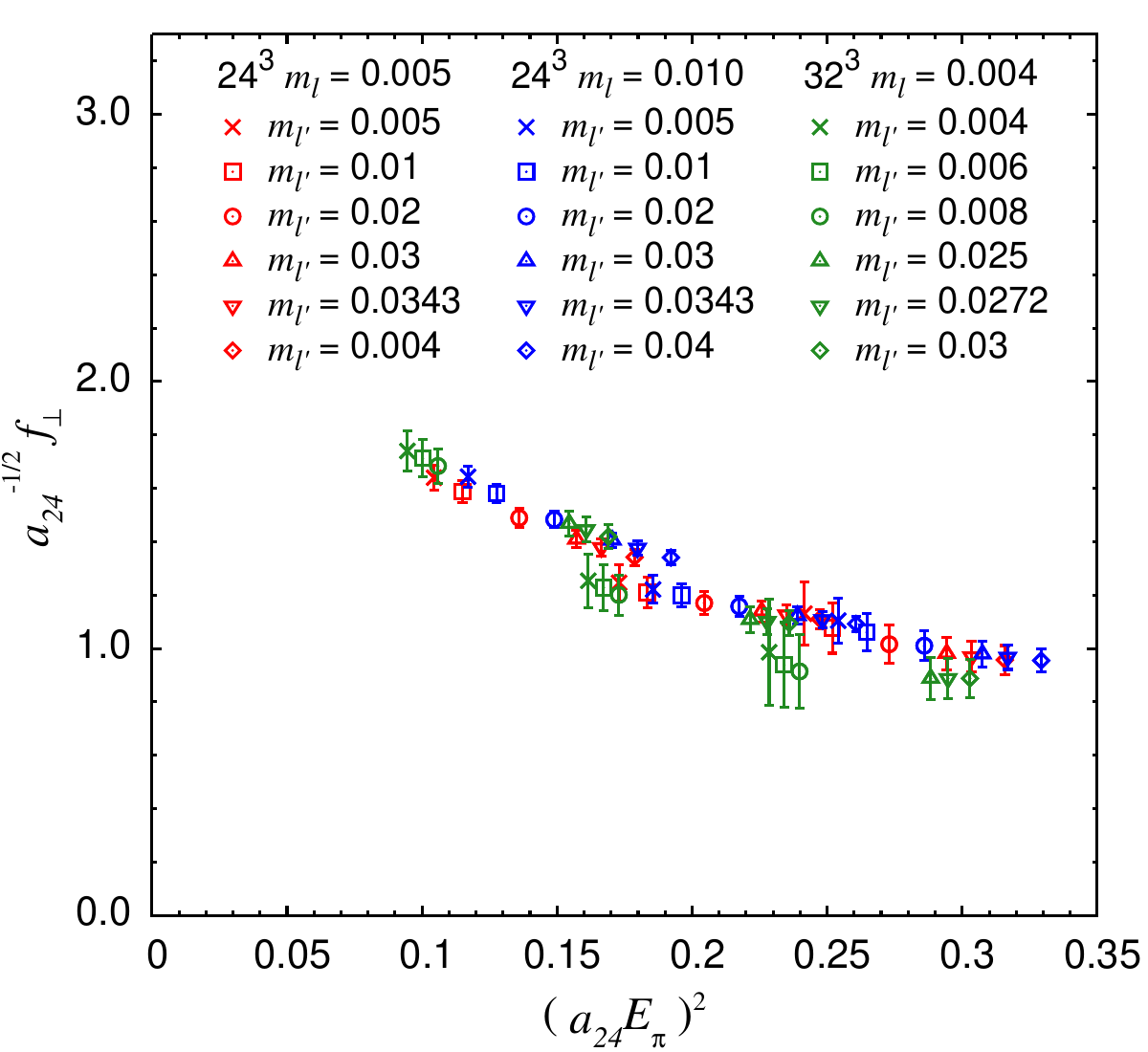}
  \caption{Form factors $f_{\parallel}$ and $ f_{\perp}$
  in coarse lattice units $a_{24}$. 
  The red, bue and green symbols  indicate  data on the 
  coarser  $am_l=0.005$, $am_l=0.01$  and finer $am_l=0.004$  ensembles,
  respectively.
Different symbols indicate different partially quenched masses.}
   \label{fig:form}
\end{figure}
After multiplying the results for $R_{3,0}^{B\to \pi}$ and $R_{3,i}^{B\to \pi}$ 
with the renormalization factor $Z_V^{bl}$,
we obtain the form factors $f_{\parallel}$ and $ f_{\perp}$ shown in Fig.~\ref{fig:form}.
(For now the $\rho$-factor is set to unity.)
After adding the $\rho$-factors and $\mathcal{O}(a)$-improvement, 
we will extrapolate to the physical quark masses and continuum and 
interpolate in $E_\pi^2$ using chiral perturbation theory.
Calculations on the remaining finer ensembles are underway.

We will then extrapolate our synthetic form-factor data over the full kinematic range (down to $q^2 = 0$)
using the model-independent $z$-parameterization~\cite{Bourrely:1980gp,Boyd:1994tt,Arnesen:2005ez,Bourrely:2008za}.
Our result will provide an important independent check on existing calculations 
using staggered light quarks.

\section{Acknowledgments}
We thank our collaborators in the RBC and UKQCD Collaborations
for helpful discussions.
Computations for this work were mainly performed on resources provided
by the USQCD Collaboration, funded by the Office of Science of the U.S. Department of
Energy, as well as computers at BNL and Columbia University.
T.K  is partially supported by JSPS Grants-in-Aid (No. 22-7653).
O.W. acknowledges support at Boston University by the U.S. DOE grant
DE-FC02-06ER41440.
BNL is operated by Brookhaven Science Associates, LLC under Contract 
No. DE-AC02-98CH10886 with the U.S. Department of Energy.  
Fermilab is operated by Fermi Research Alliance, LLC, under Contract 
No. DE-AC02-07CH11359 with the U.S. Department of Energy.

{\small
\bibliographystyle{apsrev4-1}
\bibliography{lattice2012_bibtex}
}

\end{document}